\documentclass[traditabstract]{aa}
\usepackage{graphicx}
\usepackage{txfonts}
\usepackage{color}

\begin{document}

\title{Optical circular polarization in quasars\thanks{Based on
observations made with ESO Telescopes at the La Silla Observatory
(Chile). ESO program ID: 79.A-0625(B)}}

\author{D. Hutsem\'ekers\inst{1,}\thanks{Ma{\^\i}tre de Recherches au F.R.S.-FNRS}
   \and B. Borguet\inst{2}
   \and D. Sluse\inst{3} 
   \and R. Cabanac\inst{4} 
   \and H. Lamy\inst{5} 
}

\institute{Institut d'Astrophysique et de G\'eophysique,
           Universit\'e de Li\`ege, All\'ee du 6 Ao\^ut 17, B5c, B-4000
           Li\`ege, Belgium
      \and Department of Physics, Virginia Polytechnic and State University, 
           Blacksburg, VA 24061, USA
      \and Zentrum f\"{u}r Astronomie der 
           Universit\"{a}t Heidelberg,  M\"{o}nchhofstr.\ 12-14, 
           69120 Heidelberg, Germany
      \and Laboratoire d'Astrophysique de Toulouse-Tarbes, Universit\'e
           de Toulouse, 57 avenue d'Azereix, 65000 Tarbes, France
      \and Institut Belge d'A\'eronomie Spatiale, Avenue Circulaire 3, 
           1180 Bruxelles, Belgium
}
\date{Received ; accepted: }

   

\abstract{
We present new optical circular polarization measurements with typical
uncertainties $< 0.1\%$ for a sample of 21 quasars.  All but two
objects have null circular polarization. We use this result to
constrain the polarization due to photon-pseudoscalar mixing along the
line of sight. We detect significant ($> 3\sigma$) circular
polarization in two blazars with high linear polarization and discuss
the implications of this result for quasar physics. In particular, the
recorded polarization degrees may be indicative of magnetic fields as
strong as 1~kG or a significant contribution of inverse Compton
scattering to the optical continuum.  }

   
\maketitle

%

\section{Introduction}
\label{sec:intro}

To interpret the large-scale alignments of quasar optical polarization
vectors observed at redshifts $z\sim$~1 (Hutsem\'ekers \cite{HUT98};
Hutsem\'ekers and Lamy \cite{HUT01}; Hutsem\'ekers et
al. \cite{HUT05}) polarization induced by photon-pseudoscalar mixing
along the line of sight has been invoked (Hutsem\'ekers \cite{HUT98};
Jain et al. \cite{JAI02}).  Photon-pseudoscalar mixing generates
dichroism and birefringence, the latter transforming linear
polarization into circular polarization and vice-versa along the line
of sight.  If photon-pseudoscalar mixing produces the linear
polarization needed to explain the observed alignments, a comparable
amount of circular polarization would be expected (Raffelt and
Stodolsky \cite{RAF88}; Jain et al. \cite{JAI02}; Das et
al. \cite{DAS04}; Gnedin et al. \cite{GNE07}; Hutsem\'ekers et
al. \cite{HUT08}; Payez et al. \cite{PAY08}).  Hence, we present
accurate circular polarization measurements for a sample of quasars
whose polarization vectors are coherently oriented.

The optical circular polarization of quasars has rarely been measured.
Our new observations, data reduction, and a compilation of published
measurements are presented in Sect.~\ref{sec:data}. Implications for
the photon-pseudoscalar mixing mechanism are discussed in
Sect.~\ref{sec:discuss1}. The detection of significant circular
polarization in two objects and its consequence for quasar physics are
presented in Sect.~\ref{sec:discuss2}.

\section{Observations and data reduction}
\label{sec:data}

The observations were carried out on April 18-20, 2007 at the European
Southern Observatory (ESO, La Silla) using the 3.6m telescope equipped
with the ESO Faint Object Spectrograph and Camera EFOSC2. Circular
polarization was measured using a super-achromatic quarter-wave
($\lambda$/4) retarder plate (QWP), which transforms the circular
polarization into linear polarization, and a Wollaston prism, which
splits the linearly polarized beam into two orthogonally polarized
images of the object (Saviane et al. \cite{SAV07}). The CCD was used
in unbinned mode, which corresponds to a scale of 0.157$\arcsec$/pixel
on the sky.  All measurements were performed through a Bessel V filter
(V$\#$641; central wavelength: 5476 \AA; FWHM: 1132 \AA).

At least one pair of exposures with the QWP rotated to the angles $-45
\degr$ and $+45 \degr$ was secured for each target. Frames were
dark-subtracted and flat-fielded.  The circular polarization $p_{\rm
circ}$, i.e., the normalized Stokes $V/I$ parameter, was extracted
from each pair of frames using a procedure used to measure the
normalized Stokes $Q/I$ and $U/I$ parameters and described in Lamy and
Hutsem\'ekers (\cite{LAM99}) and Sluse et al. (\cite{SLU05}).  Errors
were estimated from the photon noise. Seeing was typically
around~1$\arcsec$. Owing to the variable atmospheric extinction (thin
to thick cirrus), some exposures had to be repeated to reach a
sufficient signal-to-noise ratio.

The performances of the instrument were checked during our run and
during the setup night (April 17) using an unpolarized standard star
and a star with high and slowly variable circular polarization,
LP~790$-$20 (West \cite{WES89}; Jordan and Friedrich
\cite{JOR02}). The results, discussed in Saviane et
al. (\cite{SAV07}), demonstrated the quality of the instrumental
setup.  LP~790$-$20 was also used to fix the sign of the circular
polarization, i.e., $p_{\rm circ} > 0$ when the electric vector
rotates counter-clockwise as seen by an observer facing the object.

To evaluate the cross-talk between linear and circular polarization,
we measured the circular polarization of linearly polarized
stars. These observations were repeated several times during our
observing run. Hilt~652 was observed during the setup night. The
results are given in Table~\ref{tab:datastd} together with the
published linear polarization (i.e. the polarization degree $p_{\rm
lin}$ and the polarization position angle $\theta_{\rm
lin}$). Uncertainties are smaller than in Saviane et
al.~(\cite{SAV07}) because of the availability of repeated
observations.  Although the objects are highly linearly polarized, we
measure a null circular polarization. Combining the data of Hilt~652
and Ve~6$-$23, which have similar polarization angles, we derive the
3$\sigma$ upper limit to the circular polarization due to cross-talk
in the V filter $|p_{\rm circ} / p_{\rm lin}| \lesssim$
0.0075.

Our new measurements of quasar circular polarization are reported in
Table~\ref{tab:dataqso} with 1$\sigma$ photon-noise errors.  The
targets are extracted from the sample of 355 polarized quasars defined
in Hutsem\'ekers et al.~(\cite{HUT05}), as well as their B1950
names/coordinates, their redshift $z$, and their linear polarization
degree and angle, $p_{\rm lin}$ and $\theta_{\rm lin}$.

A compilation of other measurements of quasar optical circular
polarization is given in Table~\ref{tab:prevqso}. Unless indicated
otherwise, these measurements were obtained in white light, i.e., in
the 3200--8800~\AA\ or 4000--8800~\AA\ spectral ranges, which roughly
correspond to an effective wavelength of 6000~\AA. When several
estimates of either linear or circular polarization are available,
only the value with the smallest uncertainty is considered.  BL~Lac
objects, similar in many respects to highly polarized quasars (HPQs)
(e.g., Scarpa and Falomo \cite{SCA97}; Fan et al. \cite{FAN08}), are
included.  Both BL Lac and HPQs belong to the blazar sub-group of
active galactic nuclei (AGN).  For BL~Lac objects, the polarization is
often strongly variable.  We then adopt the circular polarization with
the smallest uncertainty and, when quasi-simultaneous observations are
available, the value of the linear polarization obtained as close as
possible in time.  Otherwise we select a representative value of the
linear polarization from the survey of Impey and Tapia~(\cite{IMP90}).

\begin{table}[t]
\caption{The circular polarization of linearly polarized standard stars}
\label{tab:datastd}
\begin{tabular}{lccr}\hline\hline \\[-0.10in]
Object &  $p_{\rm lin}$ (\%) &  $\theta_{\rm lin}$ ($\degr$) & $p_{\rm circ}$ (\%) \\ 
\hline \\[-0.10in]
Hilt 652   & 6.25  $\pm$ 0.03 & 179.2  $\pm$ 0.2$^{a}$ &     0.003 $\pm$ 0.021 \\
Ve~6$-$23  & 8.26  $\pm$ 0.05 & 171.6  $\pm$ 0.2$^{a}$ &  $-$0.050 $\pm$ 0.035 \\
HD155197   & 4.38  $\pm$ 0.03 & 103.2  $\pm$ 0.2$^{b}$ &     0.033 $\pm$ 0.025 \\  
\hline\\[-0.2cm]
\end{tabular}\\
\tiny{Note: All these polarization measurements were obtained
in the V filter. References for linear polarization: (a)~Fossati et
al. \cite{FOS07}; (b)~Turnshek et al. \cite{TUR90}. }
\end{table}

\begin{table}[t]
\caption{New circular polarization measurements of quasars}
\label{tab:dataqso}
\begin{tabular}{lcccr}\hline\hline\\[-0.10in]
Object &  $z$  &  $p_{\rm lin}$ (\%) &  $\theta_{\rm lin}$ ($\degr$) &  $p_{\rm circ}$ (\%)\\ 
\hline\\[-0.10in]
\hspace*{-1mm}1120$+$019 & \hspace*{-1mm}1.465 &   1.95 $\pm$ 0.27 &    9 $\pm$  4$^{c}$ &   $-$0.02 $\pm$ 0.05 \\ 
\hspace*{-1mm}1124$-$186 & \hspace*{-1mm}1.048 &  11.68 $\pm$ 0.36 &   37 $\pm$  1$^{g}$ &   $-$0.04 $\pm$ 0.08 \\ 
\hspace*{-1mm}1127$-$145 & \hspace*{-1mm}1.187 &   1.30 $\pm$ 0.40~[w] &   23 $\pm$ 10$^{a}$ &   $-$0.05 $\pm$ 0.05 \\ 
\hspace*{-1mm}1157$+$014 & \hspace*{-1mm}1.990 &   0.76 $\pm$ 0.18 &   39 $\pm$  7$^{f}$ &   $-$0.10 $\pm$ 0.08 \\ 
\hspace*{-1mm}1205$+$146 & \hspace*{-1mm}1.640 &   0.83 $\pm$ 0.18 &  161 $\pm$  6$^{f}$ &   $-$0.10 $\pm$ 0.09 \\ 
\hspace*{-1mm}1212$+$147 & \hspace*{-1mm}1.621 &   1.45 $\pm$ 0.30 &   24 $\pm$  6$^{c}$ &      0.15 $\pm$ 0.09 \\ 
\hspace*{-1mm}1215$-$002$^{\star}$ & \hspace*{-1mm}0.420 &  23.94 $\pm$ 0.70 &   91 $\pm$  1$^{g}$ &   $-$0.42 $\pm$ 0.40 \\
\hspace*{-1mm}1216$-$010 & \hspace*{-1mm}0.415 &  11.20 $\pm$ 0.17 &  100 $\pm$  1$^{g}$ &   $-$0.01 $\pm$ 0.07 \\ 
\hspace*{-1mm}1222$+$228 & \hspace*{-1mm}2.058 &   0.92 $\pm$ 0.14 &  169 $\pm$  4$^{g}$ &      0.01 $\pm$ 0.10 \\ 
\hspace*{-1mm}1244$-$255 & \hspace*{-1mm}0.633 &   8.40 $\pm$ 0.20~[w] &  110 $\pm$  1$^{a}$ &   $-$0.23 $\pm$ 0.20 \\ 
\hspace*{-1mm}1246$-$057 & \hspace*{-1mm}2.236 &   1.96 $\pm$ 0.18~[w] &  149 $\pm$  3$^{e}$ &      0.01 $\pm$ 0.03 \\ 
\hspace*{-1mm}1254$+$047 & \hspace*{-1mm}1.024 &   1.22 $\pm$ 0.15~[w] &  165 $\pm$  3$^{b}$ &   $-$0.02 $\pm$ 0.04 \\ 
\hspace*{-1mm}1256$-$229$^{\star}$ & \hspace*{-1mm}0.481 &  22.32 $\pm$ 0.15 &  157 $\pm$  1$^{g}$ & 0.18 $\pm$ 0.04 \\ 
\hspace*{-1mm}1309$-$056 & \hspace*{-1mm}2.212 &   0.78 $\pm$ 0.28 &  179 $\pm$ 11$^{c}$ &   $-$0.08 $\pm$ 0.06 \\ 
\hspace*{-1mm}1331$-$011 & \hspace*{-1mm}1.867 &   1.88 $\pm$ 0.31 &   29 $\pm$  5$^{c}$ &   $-$0.04 $\pm$ 0.06 \\ 
\hspace*{-1mm}1339$-$180 & \hspace*{-1mm}2.210 &   0.83 $\pm$ 0.15 &   20 $\pm$  5$^{g}$ &   $-$0.01 $\pm$ 0.07 \\ 
\hspace*{-1mm}1416$-$129 & \hspace*{-1mm}0.129 &   1.63 $\pm$ 0.15~[w] &   44 $\pm$  3$^{b}$ &      0.05 $\pm$ 0.06 \\ 
\hspace*{-1mm}1429$-$008 & \hspace*{-1mm}2.084 &   1.00 $\pm$ 0.29 &    9 $\pm$  9$^{c}$ &      0.02 $\pm$ 0.08 \\ 
\hspace*{-1mm}2121$+$050 & \hspace*{-1mm}1.878 &  10.70 $\pm$ 2.90~[w] &   68 $\pm$  6$^{a}$ &      0.02 $\pm$ 0.15 \\ 
\hspace*{-1mm}2128$-$123 & \hspace*{-1mm}0.501 &   1.90 $\pm$ 0.40~[w] &   64 $\pm$  6$^{d}$ &   $-$0.04 $\pm$ 0.03 \\ 
\hspace*{-1mm}2155$-$152 & \hspace*{-1mm}0.672 &  22.60 $\pm$ 1.10~[w] &    7 $\pm$  2$^{a}$ & $-$0.35 $\pm$ 0.10\\
\hline\\[-0.2cm]
\end{tabular}\\
\tiny{Notes: Linear and circular polarizations were measured in
the V filter except a series of linear polarization data from the
literature measured in white light and noted~[w]; (${\star}$)
1215$-$002 is classified as a BL~Lac by Collinge~et~al.~\cite{COL05};
Sbarufatti et al. \cite{SBA05} re-determined the redshift of
1256$-$229 ($z$=0.481) and considered this object as a BL Lac.
References for linear polarization: (a)~Impey \& Tapia \cite{IMP90};
(b)~Berriman et al. \cite{BER90}; (c)~Hutsem\'ekers et
al. \cite{HUT98b}; (d)~Visvanathan \& Wills \cite{VIS98}; (e)~Schmidt
\& Hines \cite{SCH99}; (f)~Lamy \& Hutsem\'ekers \cite{LAM00};
(g)~Sluse et al. \cite{SLU05}. }
\end{table}

\begin{table}[t]
\caption{Previous circular polarization measurements of quasars and BL~Lac objects}
\label{tab:prevqso}
\begin{tabular}{lcccr}\hline\hline \\[-0.10in]
Object &  $z$  &  $p_{\rm lin}$ (\%) &  $\theta_{\rm lin}$ ($\degr$) &  $p_{\rm circ}$ (\%) \\ 
\hline \\[-0.10in]
\hspace*{-1mm}0237$-$233 & 2.223 &   0.25 $\pm$ 0.29     &   $-$ \      $^{d}$ & $-$0.06 $\pm$ 0.08$^{a}$ \\
\hspace*{-1mm}0955$+$326 & 0.533 &   0.18 $\pm$ 0.24     &   $-$ \      $^{c}$ &    0.06 $\pm$ 0.08$^{a}$ \\
\hspace*{-1mm}1127$-$145 & 1.187 &   1.30 $\pm$ 0.40     &  23 $\pm$ 10$^{e}$ &    0.32 $\pm$ 0.20$^{a}$ \\ 
\hspace*{-1mm}1156$+$295 & 0.729 &   2.68 $\pm$ 0.41     & 114 $\pm$  4$^{f}$ &    0.12 $\pm$ 0.14$^{b}$ \\
\hspace*{-1mm}1222$+$228 & 2.058 &   1.09 $\pm$ 0.16~[u] & 167 $\pm$  4$^{i}$ &    0.23 $\pm$ 1.80$^{i}$ \\
\hspace*{-1mm}1226$+$023 & 0.158 &   0.25 $\pm$ 0.04     &  58 $\pm$  4$^{c}$ & $-$0.01 $\pm$ 0.02$^{g}$ \\
\hspace*{-1mm}1253$-$055 & 0.536 &   9.00 $\pm$ 0.40     &  67 $\pm$  1$^{e}$ &    0.09 $\pm$ 0.07$^{a}$ \\
\hspace*{-1mm}1308$+$326 & 0.997 &  12.10 $\pm$ 1.50     &  68 $\pm$  3$^{e}$ & $-$0.08 $\pm$ 0.17$^{b}$ \\
\hspace*{-1mm}1634$+$706 & 1.334 &   0.24 $\pm$ 0.07~[u] &   4 $\pm$  8$^{i}$ & $-$0.05 $\pm$ 0.09$^{i}$ \\
\hspace*{-1mm}1641$+$399 & 0.594 &   4.00 $\pm$ 0.30     & 103 $\pm$  2$^{e}$ & $-$0.05 $\pm$ 0.23$^{b}$ \\
\hspace*{-1mm}2230$+$114 & 1.037 &   7.30 $\pm$ 0.30     & 118 $\pm$  1$^{e}$ & $-$0.05 $\pm$ 0.17$^{a}$ \\
\hspace*{-1mm}2302$+$029 & 1.044 &   0.66 $\pm$ 0.12~[u] & 136 $\pm$  5$^{i}$ & $-$0.39 $\pm$ 0.16$^{i}$ \\
\hspace*{-1mm}{\it 0138$-$097} & 0.733 &  3.60 $\pm$ 1.50 & 168 $\pm$ 11$^{e}$ & 0.25 $\pm$ 0.35$^{k}$ \\
\hspace*{-1mm}{\it 0219$+$428} & 0.444 & 26.11 $\pm$ 0.19 &   3 $\pm$  1$^{k}$ & 0.16 $\pm$ 0.05$^{k}$ \\
\hspace*{-1mm}{\it 0422$+$004} & 0.310 & 10.29 $\pm$ 0.23 & 179 $\pm$  1$^{g}$ &    0.14 $\pm$ 0.07$^{g}$ \\
\hspace*{-1mm}{\it 0735$+$178} & 0.424 & 11.69 $\pm$ 0.22 & 123 $\pm$  1$^{k}$ & 0.03 $\pm$ 0.05$^{k}$ \\
\hspace*{-1mm}{\it 0823$-$223} & 0.910 & 14.39 $\pm$ 0.16 &  11 $\pm$  1$^{k}$ & 0.16 $\pm$ 0.08$^{k}$ \\
\hspace*{-1mm}{\it 0851$+$202} & 0.306 & 10.80 $\pm$ 0.30 & 156 $\pm$  1$^{e}$ & $-$0.01 $\pm$ 0.02$^{a}$ \\
\hspace*{-1mm}{\it 1101$+$384} & 0.031 &  2.59 $\pm$ 0.11 &  10 $\pm$  1$^{h}$ & 0.02 $\pm$ 0.03$^{h}$ \\
\hspace*{-1mm}{\it 2155$-$304} & 0.116 &  4.12 $\pm$ 0.25 &  93 $\pm$  2$^{j}$ & $-$0.02 $\pm$ 0.02$^{j}$ \\
\hspace*{-1mm}{\it 2200$+$420} & 0.068 &  4.90 $\pm$ 0.40 & 147 $\pm$  2$^{e}$ & $-$0.07 $\pm$ 0.19$^{a}$ \\
\hline\\[-0.2cm]
\end{tabular}\\
\tiny{Notes: All but a few polarization measurements were
obtained in white light; the multi-color measurements in references
(h), (j) and (k) were averaged; [u] refers to linear and circular
polarization measurements averaged over the 2200--3200 \AA\
ultraviolet wavelength band; the circular polarization of objects
2200$+$420 and 2230$+$114 was obtained in the 4000--6000~\AA\ and
3500--5200~\AA\ bands respectively; italicized names indicate objects
classified as BL~Lac in V\'eron-Cetty \& V\'eron
\cite{VER06}. References for linear and circular polarization:
(a)~Landstreet \& Angel \cite{LAN72}; (b)~Moore \& Stockman
\cite{MOO81}; (c)~Stockman et al. \cite{STO84}; (d)~Moore \& Stockman
\cite{MOO84}; (e)~Impey \& Tapia \cite{IMP90}; (f)~Wills et
al. \cite{WIL92}; (g)~Valtaoja et al. \cite{VAL93}; (h)~Takalo \&
Sillanp\"a\"a \cite{TAK93}; (i)~Impey et al. \cite{IMP95}; (j)~Tommasi
et al. \cite{TOM01a}; (k)~Tommasi et al. \cite{TOM01b}.  }
\end{table}

\section{Discussion}
\label{sec:discuss}

The measurements reported in Tables~\ref{tab:dataqso}
and~\ref{tab:prevqso} show that all quasars and BL Lac objects have
null circular polarization ($<$~3~$\sigma$) except two HPQs,
1256$-$229 and 2155$-$152, and one highly polarized BL Lac object,
0219$+$428.

We first discuss the constraints provided by the majority of null
detections on the photon-pseudoscalar mixing mechanism, and then the
consequences of the three detections for blazar physics.

\subsection{Constraints on photon-pseudoscalar mixing}
\label{sec:discuss1}

Quasars with right ascension between 11$^{\rm h}$20$^{\rm m}$ and
14$^{\rm h}$30$^{\rm m}$ belong to the region of alignment A1 defined
in Hutsem\'ekers~(\cite{HUT98}). In this region of the sky, quasars
with $1 < z < 2.3$ have their polarization angle preferentially in the
range [146$\degr$--226$\degr$] (modulo 180$\degr$), while quasars with
$0 < z < 1$ have their polarization angle preferentially in the range
[30$\degr$--120$\degr$]. Assuming that the quasar intrinsic
polarization vectors are randomly oriented, the addition of a small
systematic linear polarization $\Delta p_{\rm lin} \simeq 0.5 \%$ at a
fixed position angle can account for the observed alignments
(Hutsem\'ekers et al. \cite{HUT08}; Appendix~\ref{sec:apa}).  If
photon-pseudoscalar mixing is responsible for this extra linear
polarization, one expects, on average, that $| p_{\rm circ}| \simeq
\Delta p_{\rm lin} \simeq 0.5 \%$ (Appendix~\ref{sec:apb}). Because
the light from most quasars is intrinsically linearly polarized to
some extent and not circularly polarized, limits on any additional
polarization from interactions along the line of sight cannot be
derived from the measurement of the linear polarization degree, while,
on the other hand, useful constraints can be derived from the
measurement of circular polarization.

Most of the thirteen quasars with $z > 1$ located in region A1 were
found to have $|p_{\rm circ}| \lesssim 0.25 \% $ (3$\sigma$ upper
limit), which is definitely smaller than the expected value. Averaging
over the thirteen objects, we infer that $\langle |p_{\rm circ}|
\rangle = 0.035 \pm 0.016 \%$ after neglecting the sign, from which a
stringent 3$\sigma$ upper limit on the circular polarization of
$\langle |p_{\rm circ}| \rangle \leq 0.05\%$ can be derived. This
limit is one order of magnitude smaller than the expected value $|
p_{\rm circ}| \simeq 0.5 \%$. A similar result is obtained for the
nine objects at $z < 1$ in that region.

This result rules out the interpretation of the observed alignments in
terms of photon-pseudoscalar mixing, at least in its simplest
formulation. A more complex treatment of the photon-pseudoscalar
interaction is thus required to account for the observations (Payez et
al. \cite{PAY10a,PAY10b}).

\subsection{Detection of optical circular polarization and 
implication for blazar physics}
\label{sec:discuss2}

Circular polarization is detected at the 3$\sigma$ level in two HPQs:
1256$-$229 and 2155$-$152 (Table~\ref{tab:dataqso}).  On April 21, we
had the opportunity to re-measure the linear polarization of these
objects in the V filter, after replacing the quarter-wave plate by the
half-wave plate (HWP) (cf. Saviane et al. \cite{SAV07}). Four
exposures with the HWP rotated to 0$\degr$, 22.5$\degr$, 45$\degr$,
and 67.5$\degr$ were secured and reduced in the standard way
(e.g. Sluse et al. \cite{SLU05}).  The results are reported in
Table~\ref{tab:circqso}, together with the circular polarization
measurements from Table~\ref{tab:dataqso}. Although the optical linear
polarization of these quasars is high, the circular polarization we
measured at the same epoch is above the 3$\sigma$ upper limit on the
circular polarization generated by the instrumental cross-talk
(Sect.~\ref{sec:data}).

In Table~\ref{tab:circqso}, we also summarize the main polarization
properties of these objects, including measurements at radio
wavelengths. For completeness, we include the BL Lac object for which
circular polarization was found to be significant after averaging the
UBVRI measurements (Table~\ref{tab:prevqso}).  As far as we know,
these are the only 3$\sigma$ detections of optical circular
polarization in quasars, in addition to those reported by Wagner and
Mannheim (\cite{WAG01}) for 3C279\footnote{Wagner et
al. (\cite{WAG00}) reported that $p_{\rm circ} = 0.25 \pm 0.03 \%$,
while Wagner and Mannheim (\cite{WAG01}) reported $p_{\rm circ} =
0.45 \pm 0.03 \%$. These detections were considered tentative by the
authors in view of significant instrumental effects and thus not
included in Tables~\ref{tab:prevqso} and~\ref{tab:circqso}.} (=
1253$-$055).  Although variable (as commonly seen in HPQs), the
optical linear polarization is high in all three objects suggesting
that a relation exists between linear and circular polarization.

\begin{table}[t]
\caption{Radio to optical polarization characteristics of objects 
with detected optical circular polarization}
\label{tab:circqso}
\begin{tabular}{lcccr}\hline\hline\\[-0.10in]
\multicolumn{5}{l}{1256$-$229~(PKS)  \ \ $z$=0.481}  \ \ \ HPQ, BL Lac? \ \  \\ 
\hline\\[-0.10in]
Date &  $\nu$ (Ghz) &  $p_{\rm lin}$ (\%) &  $\theta_{\rm lin}$ ($\degr$) &  $p_{\rm circ}$ (\%)\\ 
\hline\\[-0.10in]
03/2002 &  5.4 10$^5$ &  22.32 $\pm$ 0.15 & 157 $\pm$  1$^{d}$ & $-$ \ \    \\ 
04/2007 &  5.4 10$^5$ &  15.42 $\pm$ 0.16 & 163 $\pm$  1$^{j}$ & 0.18 $\pm$ 0.04$^{j}$ \\
\hline\hline\\[-0.10in]
\multicolumn{5}{l}{2155$-$152~(PKS)  \ \ $z$=0.672}  \ \ \ HPQ \ \  \\ 
\hline\\[-0.10in]
Date &  $\nu$ (Ghz) &  $p_{\rm lin}$ (\%) &  $\theta_{\rm lin}$ ($\degr$) &  $p_{\rm circ}$ (\%)\\ 
\hline\\[-0.10in]
08/1984 &  5.4 10$^5$  &  32.70 $\pm$ 1.30 &   6 $\pm$  1$^{a}$ &  $-$ \ \     \\
04/2007 &  5.4 10$^5$  &  17.67 $\pm$ 0.49 &  51 $\pm$  1$^{j}$ & $-$0.35 $\pm$ 0.10$^{j}$\\
09/2007 &        85.6 &  11.04 $\pm$ 0.53 &  52 $\pm$  1$^{i}$ &  $<$ 0.93$^{i}$    \\
03/2003 &        15.4 &  3.66  & $-$  \ $^{f}$ &  $<$ 0.34$^{f}$    \\
1979-1999 &       8.0 &  4.81 (mean)  & $-$  \ $^{h}$ &  $-$ \ \     \\
          &           &  15.3 (max) & $-$  \ $^{h}$ &  $-$ \ \     \\
\hline\hline\\[-0.10in]
\multicolumn{5}{l}{0219$+$428~(3C66A)  \ \ $z$=0.444}  \ \ \ BL Lac \ \  \\ 
\hline\\[-0.10in]
Date &  $\nu$ (Ghz) &  $p_{\rm lin}$ (\%) &  $\theta_{\rm lin}$ ($\degr$) &  $p_{\rm circ}$ (\%)\\ 
\hline\\[-0.10in]
01/1992 &  5 10$^5$~[w] &  31.07 $\pm$ 0.31 &  39 $\pm$  1$^{b}$ & 0.76 $\pm$ 0.10$^{b}$\\
12/1999 &  5 10$^5$~[w] &  26.11 $\pm$ 0.19 &  3 $\pm$  1$^{c}$ & 0.16 $\pm$ 0.05$^{c}$\\
11/2008 &      85.6 &   4.36 $\pm$ 0.54 &  3 $\pm$  4$^{i}$ &  $<$ 0.73$^{i}$    \\
12/1996 &       5.0 &  2.9  & $-$  \ $^{e}$ &  $<$ 0.20$^{e}$    \\
1974-1999 &     8.0 &  2.96 (mean)  & $-$  \ $^{g}$ &  $-$ \ \     \\
\hline\\[-0.2in]
\end{tabular}
$ $\\[0.1in]
\tiny{Notes: Multi-color measurements in references (b) and (c)
were averaged; upper limits are given at 3$\sigma$. References for
linear or circular polarization: (a)~Brindle et al. \cite{BRI86};
(b)~Takalo \& Sillanp\"a\"a \cite{TAK93}; (c)~Tommasi et
al. \cite{TOM01b}; (d)~Sluse et al. \cite{SLU05}; (e)~Homan et
al. \cite{HOM01}; (f)~Homan and Lister \cite{HOM06}; (g)~Fan et
al. \cite{FAN06}; (h)~Fan et al. \cite{FAN08}; (i)~Agudo et
al. \cite{AGU10}; (j)~this work. }
\end{table}

Radio circular polarization has been detected in a small number of
blazars with typical values of a few tenths of a percent (Weiler and
de Pater \cite{WEI83}; Rayner et al. \cite{RAY00}; Homan et
al. \cite{HOM01}; Homan and Lister \cite{HOM06}; Vistrishchak et
al. \cite{VIS08}). Although the origin of the radio circular
polarization is not yet understood, two main mechanisms of production
have been proposed: intrinsic circular polarization of the
relativistically beamed synchrotron radiation (which also produces the
radio linear polarization) and Faraday conversion of linear to
circular polarization (e.g. Wardle and Homan \cite{WAR03}). Since
beamed synchrotron radiation can also explain the high optical linear
polarization observed in HPQs and contribute significantly to the
optical continuum (Impey and Tapia~\cite{IMP90}; Wills et
al. \cite{WIL92}), a similar origin to both the optical and the radio
circular polarizations appears likely although they most probably
arise from different regions.  Since Faraday conversion is inefficient
at visible wavelengths, the detected optical circular polarization
should be caused by synchrotron emission.  Since intrinsic circular
polarization is not produced in a positron-electron plasma, this
mechanism requires the predominance of a proton-electron plasma, as
already suggested by circular polarization measurements obtained at
millimeter wavelengths (Agudo et al. \cite{AGU10}).  Furthermore, if
circular polarization is intrinsic, a correlation between the linear
and the circular polarization degrees is expected, the high recorded
values indicating a rather homogeneous magnetic field whose strength
should be of the order of 1~kG (e.g. Valtaoja et
al. \cite{VAL93}). This is much higher than usually assumed in quasar
jets, and can only occur in small regions close to the quasar core
(Wardle and Homan \cite{WAR03}; Silant'ev et al. \cite{SIL09};
Piotrovich et al. \cite{PIO10}). On the other hand, the optical
continuum could predominantly arise from inverse Compton scattering of
radio synchrotron radiation, a mechanism that preserves the circular
polarization (Sciama and Rees \cite{SCI67}). This would require a
significant circular polarization at radio wavelengths, which is
apparently not observed (Table~\ref{tab:circqso}). Given the
uncertainties and the non-simultaneous observations, no firm
conclusion can be derived.  Unveiling the origin of the optical
circular polarization --even a few tenths of a percent-- thus appears
challenging (see also Rieger and Mannheim \cite{RIE05}). A clearer
understanding would require simultaneous observations at radio and
optical wavelengths.

\section{Conclusions}
\label{sec:conclu}

We have reported new accurate measurements of optical circular
polarization in the V filter for a sample of 21 quasars.  For most
objects, the uncertainties are smaller than 0.1\%, and smaller than
0.05\% for six of them.

All objects have null polarization within the uncertainties except two
highly linearly polarized blazars. This has allowed us to constrain
the polarization caused by photon-pseudoscalar mixing along the line
of sight, ruling out the interpretation of the observed alignments of
quasar polarization vectors in terms of photon-pseudoscalar mixing, at
least in the framework of a simple formulation.

We also found small but significant optical circular polarization in
two blazars, providing clues about the strength of the magnetic
fields, the nature of the jets and/or the dominant emission
mechanism. Our observations demonstrate that optical circular
polarization is routinely measurable with present day high-accuracy
polarimeters.

\begin{acknowledgements}
D.H. thanks Alexandre Payez and Jean-Ren\'e Cudell for useful
discussions. A fellowship from the Alexander von Humboldt Foundation
to D.S. is gratefully acknowledged. This research has made use
of data originally from the University of Michigan Radio Astronomy
Observatory, which has been supported by the University of Michigan
and the National Science Foundation.
\end{acknowledgements}

\appendix 
\section{The determination of  $\Delta p_{\rm lin}$}
\label{sec:apa}

Although partly discussed in previous papers (e.g. Hutsem\'ekers et
al. \cite{HUT05}), we provide some details of the simulations
performed to estimate $\Delta p_{\rm lin}$, the small additional
linear polarization needed to reproduce the observed alignments of
quasar polarization vectors. These simulations extend those discussed
in Hutsem\'ekers and Lamy (\cite{HUT01}), accounting for the
measurement errors.

We first modeled the distribution of the debiased polarization degree
in the full quasar sample (from Hutsem\'ekers et al. \cite{HUT05},
including objects with $p < 0.6\%$). We found that it was reasonably
well reproduced by a half-gaussian distribution of zero mean and
unitary variance. According to this distribution, we randomly generate
80 values of the polarization degree $p$, most of them lying between
0\% and 3\%. We also generate 80 values of the polarization angle
$\theta$, uniformly distributed between 0$\degr$ and 180$\degr$. From
$p$ and $\theta$, we compute the normalized Stokes parameters $q$ and
$u$ to which we add a systematic polarization $\Delta q > 0$ (we
assume for simplicity that $\Delta u = 0$, which corresponds to add
linear polarization at $\theta = 0\degr$). We also add random noise
uniformly generated between $-0.2\%$ and $+0.2\%$, in agreement with
the uncertainties in the measurements (Hutsem\'ekers et
al. \cite{HUT05}).  Finally, from the modified $q$ and $u$ we
recompute $p$, $\theta$, and $\sigma_{\theta}$ in the usual way and
select the good quality measurements with the criteria previously
used, i.e., $p \geq 0.6\%$ and $\sigma_{\theta} \leq 14\degr$. This
leaves us with $\sim$ 60 polarization values. This is comparable to
the number of objects in the alignment region (cf. Fig.~7 of
Hutsem\'ekers et al. \cite{HUT05}, to which these simulations should
be compared).

\begin{figure}[t]
\resizebox{\hsize}{!}{\includegraphics*{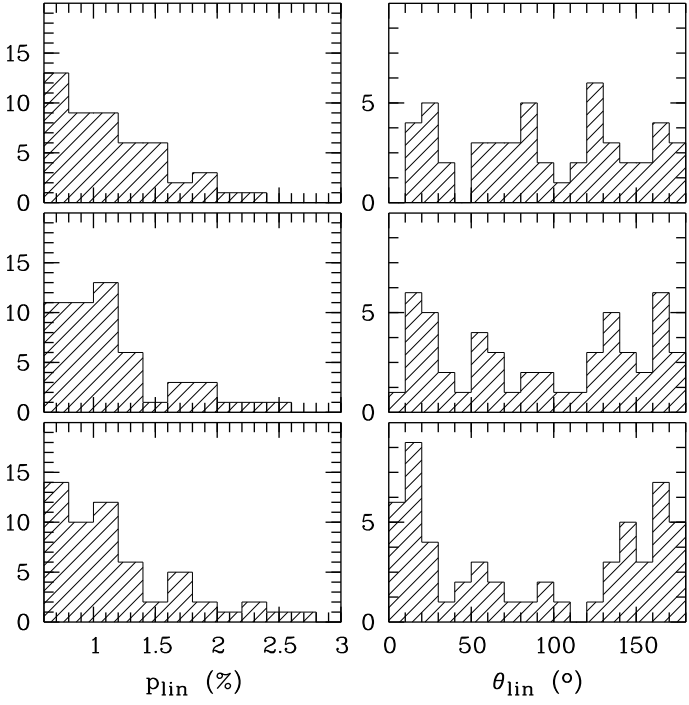}}
\caption{The effect of the addition of a small systematic polarization
$\Delta p_{\rm lin}$ on the distributions of the polarization degree
$p_{\rm lin}$ and of the polarization angle $\theta_{\rm lin}$. From
top to bottom, $\Delta p_{\rm lin}$ = 0\%, 0.25\%, and 0.5\% (see text
for details).}
\label{fig:sim}
\end{figure}

The results are illustrated in Fig.~\ref{fig:sim} for $\Delta q$ =
0\%, 0.25\%, and 0.5\% from top to bottom. We see that, in the
distribution of polarization angles, a significant deviation to
uniformity is only obtained for $\Delta p_{\rm lin}$ = $\Delta q =
0.5\%$.  At the same time, the distribution of the polarization degree
does not appear significantly modified. Since this additional
polarization at a single polarization angle is likely unrealistic,
$\Delta p_{\rm lin}$ should be seen as a lower limit, although it
cannot be much higher. Indeed, much larger values would make the
distribution of the polarization degree incompatible with the
observations (e.g. Hutsem\'ekers and Lamy \cite{HUT01}).

\section{Circular polarization due to photon-pseudoscalar mixing}
\label{sec:apb}

In the weak mixing case, photons with polarizations parallel to an
external magnetic field $\vec{B}$ that propagate through the distance
$L$ can decay into pseudoscalars with a probability
\begin{equation}
P_{\gamma a} \simeq  (\varg Bl)^{2} \, \sin^{2} (\xi/2) \, ,
\end{equation}
where $l = 2\omega/(\omega^2_p-m^2_a)$ and $\xi = L/l$, $\omega$ is
the photon frequency, $\omega_p$ the plasma frequency, $m_a$ the
pseudoscalar mass, and $\varg$ the photon-pseudoscalar coupling
constant (Raffelt and Stodolky \cite{RAF88}; Jain et
al. \cite{JAI02}).  As long as $P_{\gamma a}$ is small, the linear
polarization perpendicular to $\vec{B}$ generated by dichroism can be
approximated by $\Delta p_{\rm lin} = P_{\gamma a}$.  The mixing also
induces a polarization-dependent phase shift (retardance)
\begin{equation}
\phi_a \simeq \left( \frac{\varg Bl}{2} \right) ^{2} \, (\xi - \sin \xi )
\end{equation}
acquired by the photons during propagation, which results in
circular polarization. As noted by Raffelt and Stodolky (\cite{RAF88}),
both effects are on the order of $(\varg Bl)^{2}$.

Assuming $m_a \ll \omega_p$, we have $l \simeq 4 \times 10^{-14} \, \nu \,
n_e^{-1}$ Mpc, where $\nu$ is the frequency in GHz and $n_e$ the
electronic density in cm$^{-3}$. At optical wavelengths ($\nu =
5\times10^{5}$ GHz) and under various conditions (e.g.  $\, n_e \sim
10^{-6}$ cm$^{-3}$ and $L \sim 10$ Mpc in superclusters, or $n_e \sim
10^{-8}$ cm$^{-3}$ and $L \sim 1$ Gpc in the intergalactic medium),
$\xi = L/l \sim 500$.  With a frequency bandwidth $\Delta\nu / \nu
\sim 0.2$ and $\xi \gg 1$, we find that $\phi_a \simeq \langle \Delta
p_{\rm lin} \rangle \, \xi /2 \sim 10^{2} \, \langle \Delta p_{\rm
lin} \rangle$, where $\langle \Delta p_{\rm lin} \rangle$ represents
the average value of $\Delta p_{\rm lin}$. Similar estimates are
derived when accounting for density fluctuations (Jain et
al. \cite{JAI02}).

Adopting the convention of $u=0$ and $q>0$ for polarization vectors
parallel to $\vec{B}$, the dichroism and birefringence induced by
photon-pseudoscalar mixing modify the polarization according~to
\begin{eqnarray}
q & = &  q_{_0} -  \Delta p_{\rm lin} \, ,\nonumber \\
u & = &  u_{_0} \cos \phi_a \, , \\
\varv & = &  u_{_0} \sin \phi_a \, , \nonumber
\end{eqnarray}
where $q_{_0}$ and $u_{_0}$ are the normalized Stokes parameters
representing the initial linear polarization state and $\varv_{_0} =
0$.  Assuming that the sources are initially polarized at $p_{_0}
\simeq 2\%$ with randomly oriented polarization angles and that
$\langle \Delta p_{\rm lin} \rangle < 0.01$ (Appendix \ref{sec:apa}),
we finally obtain the average circular polarization expected to result
from the photon-pseudoscalar mixing: $\langle | \varv | \rangle \,
\sim (2/ \pi) p_{_0} \, \phi_a$, i.e., $\langle |p_{\rm circ}| \rangle
\, \sim \langle \Delta p_{\rm lin} \rangle \sim 0.5\%$. This estimate
applies to a variety of plausible situations, in agreement with the
simulations shown in Das et al. (\cite{DAS04}), Hutsem\'ekers et
al. (\cite{HUT08}), and Payez et al. (\cite{PAY08}).

\end{document}